# Genome-inspired molecular identification in organic matter via Raman spectroscopy

Yun Liu[1*†], Nicola Ferralis[1*†], L. Taras Bryndzia[2], and Jeffrey C. Grossman[1*]

[1]Department of Materials Science and Engineering, Massachusetts Institute of Technology, Cambridge, Massachusetts, United States.

[2] Shell International Exploration and Production Inc., Houston, Texas, United States.

†These authors contributed equally to this work.

**Abstract:**

Rapid, non-destructive characterization of molecular level chemistry for organic matter (OM) is experimentally challenging. Raman spectroscopy is one of the most widely used techniques for non-destructive chemical characterization, although it currently does not provide detailed identification of molecular components in OM, due to the combination of diffraction-limited spatial resolution and poor applicability of peak-fitting algorithms. Here, we develop a genome-inspired collective molecular structure fingerprinting approach, which utilizes *ab initio* calculations and data mining techniques to extract molecular level chemistry from the Raman spectra of OM. We illustrate the power of such an approach by identifying representative molecular fingerprints in OM, for which the molecular chemistry is to date inaccessible using non-destructive characterization techniques. Chemical properties such as aromatic cluster size distribution and H/C ratio can now be quantified directly using the identified molecular fingerprints. Our approach will enable non-destructive identification of chemical signatures with their correlation to the preservation of biosignatures in OM, accurate detection and quantification of environmental contamination, as well as objective assessment of OM with respect to their chemical contents.

## 1. Introduction

* Corresponding Authors: yunl@mit.edu (Yun Liu), ferralis@mit.edu (Nicola Ferralis), jcg@mit.edu (Jeffrey Grossman)

Organic matter (OM) is a typical example of a complex, highly chemically diverse geological material whose complexity reflects the inherently chemically diverse biological origin. For such material, current chemical characterization techniques provide either atomic scale identification of individual molecular structures at the cost of losing spatial resolution, or the spatial distribution of average composition at the microscale (Fig. 1). Among all chemical characterization techniques, Raman spectroscopy is unique since it is a non-destructive method capable of identifying the molecular level chemistry of materials through the identification of the vibrational signatures associated with atomic and electronic structures and specific types of chemical bonds [1-3]. However, when applied to heterogeneous organic matter, the resolution of chemistry from Raman spectroscopy is limited to the average compositional information at the micron scale [4-6], and often based on empirical and indirect correlative analyses with other techniques [4, 7-12]. The core of such limitations stems from our inability to process the large amount of information contained within the Raman spectra, which is the collective response of potentially many thousands of possible organic molecular structures [13]. For carbonaceous materials with high aromaticity, this collective response in Raman manifests as two broad bands in the 1100 – 1700 $cm^{-1}$ region (Fig. 2a), representative of molecular vibrations among different bonding configurations of carbon and hydrogen atoms. The band in the 1150-1400 $cm^{-1}$ region is called the D peak, while the one in the 1550-1600 $cm^{-1}$ region is called the G peak [13-15]. These two bands are empirically associated with the shape and the size of the aromatic structures [5, 13]. To date, the averaging across a massively large set of spectral features into single broad bands has prevented the association of any spectral feature to a particular molecular component in these materials.

Here we propose a genome-inspired molecular chemistry analysis approach that combines *ab initio* calculations with data-mining techniques to *explicitly* capture the full molecular complexity from the Raman spectrum. Unlike the top-down methods of conventional analysis algorithms that decompose broad spectral features into peaks with specific functional shapes (Fig. 2), our approach analyzes the information from a bottom-up approach, by searching for ensembles of molecular structures of which the collective Raman spectra reproduce the broad spectral features observed. Analogous to identifying genes to understand the functionality of living species, we identify "molecular fingerprints" to understand the chemistry of materials, where each "molecular fingerprint" corresponds to a single molecular structure within the ensemble. The identified ensemble of molecular fingerprints essentially serves as the "genome" of molecular chemistry for the target material, which can then be used for further analysis of materials properties and functionalities.

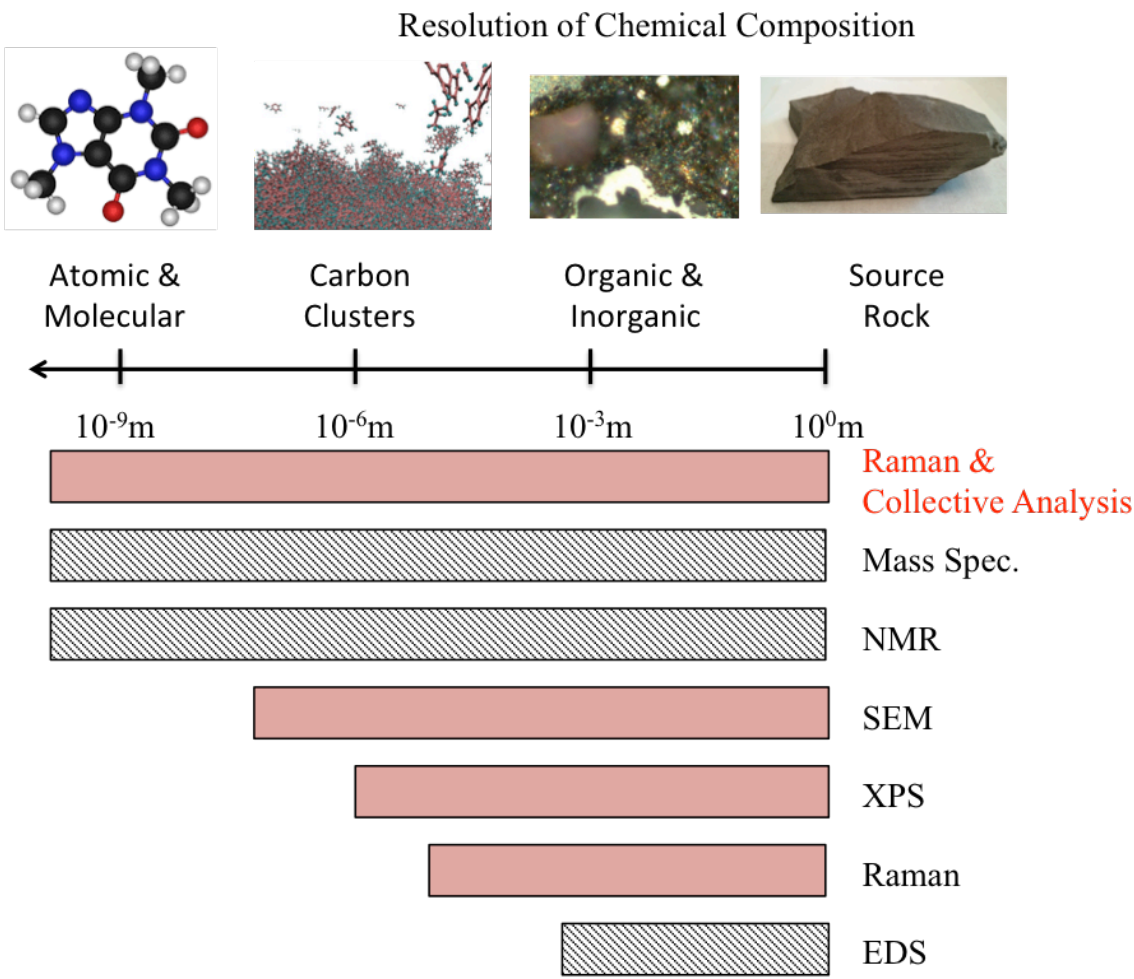

Fig. 1. Comparison of current experimental characterization techniques for organic matter with their resolution of chemistry. Techniques that can provide spatial information (original location of the content) are marked with solid red bars (and shaded bars for those which cannot, respectively). Current techniques sacrifice spatial information for resolving molecular structures. Our collective analysis approach provides not only identification of molecule structures, but also their original spatial locations. Molecular structures are visualized using the VMD software package [16].

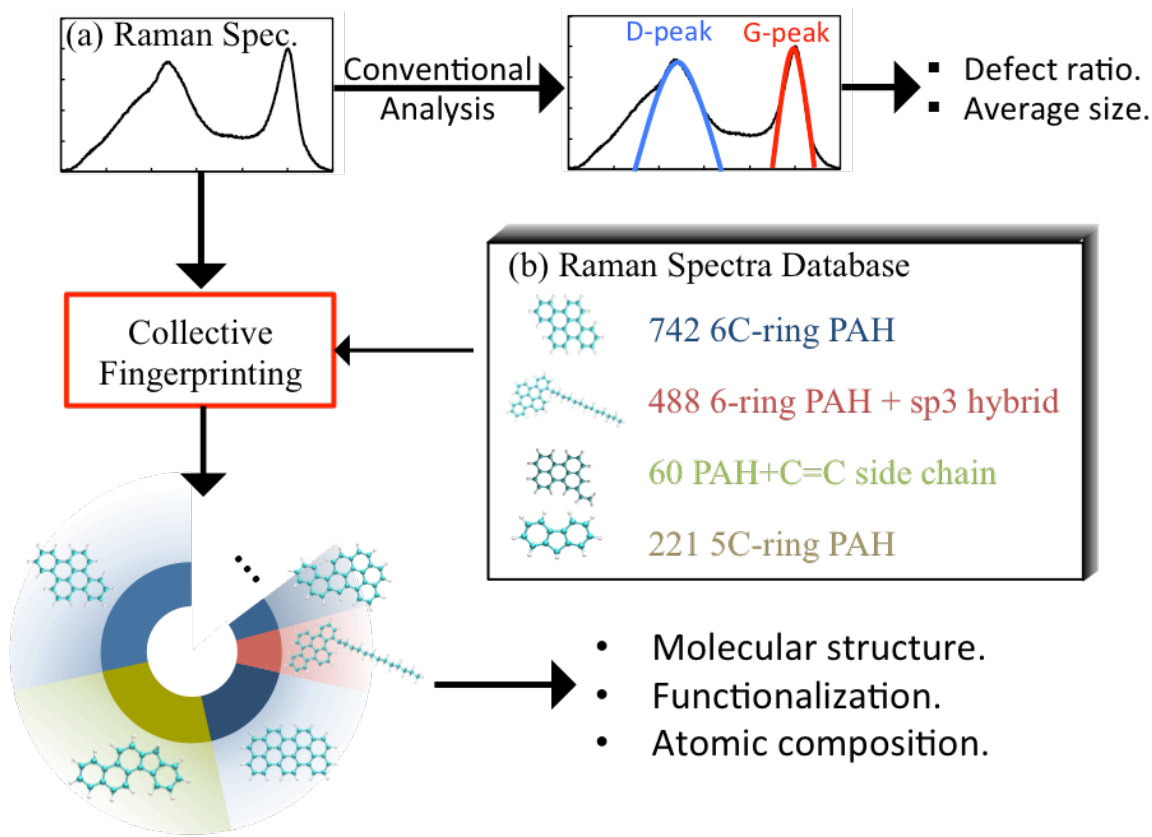

**Fig. 2.** Comparison between the conventional analysis and the collective fingerprinting approach we are taking to resolve the molecular level chemistry information from a regular Raman spectrum. (a) A typical experimental Raman spectrum of chemically heterogeneous carbonaceous materials in the 1100 – 1700 cm$^{-1}$ region. (b) List of molecular structures currently available in the "kerogen genome database" [17].

## 2. Computational and Experimental Details

### 2.1. Overview

Our approach is realized in two steps. Using *ab initio* calculations, we first build a library of computationally derived Raman spectra for a large number of molecules, the variety of which in principle covers the potential compositional phase space of the target material. Specifically for the analysis of the molecular chemistry of OM, we built a "kerogen genome database" [17] containing 1511 hydrocarbon molecules (Fig. 2b). Construction of these hydrocarbon compounds is described in section 2.2. Next, we identify molecular fingerprints using data mining techniques to determine the correlation between the experimental Raman spectrum and the Raman spectra of ensembles of molecules from the library. The ensemble with the highest correlation is most closely identified as the molecular chemistry of the target material. Specifically, our "collective fingerprinting analysis" is performed by searching for the ensemble of molecules that minimizes the normalized deviation $D$ between the experimental Raman spectrum and the collective Raman spectrum of molecules:

$$D = \frac{\sum (I_{exp} - \sum \beta_i I_i)^2}{\sum I_{exp}^2} \quad (1)$$

where $I_{exp}$ is the experimental Raman intensity at a given wavenumber, $I_i$ is the Raman intensity of the i-th molecule in the ensemble at the same wavenumber, $\beta_i$ is the linear coefficient for the i-th molecule, and the summation runs through all wavenumber points of the experimental Raman spectrum. The value of $D$ ranges between 0 and 1 [18]. From our results, Raman spectra comprised of ensembles of molecules with $D$ values below 0.1 typically represent well the main features of the experimental spectrum. However, practically, the exponential number of possible combinations with respect to the ensemble size rapidly overwhelms any attempt at a sequential search through the Raman library. To overcome this hurdle, a genetic algorithm [19] is employed to accelerate the search (see Section 2.4).

2.2. Construction of hydrocarbon molecular structures with random walk algorithm.

To ensure a systematic generation of hydrocarbon molecules with various types of aromatic and aliphatic components, we designed an algorithm that takes into account independently the aromatic size, aliphatic ratio, and the position where the aliphatic chains are attached to the aromatic flake. The aromatic flakes are generated from a random walk on a two dimensional graphene lattice. The aliphatic chains are generated at a random length between [1, $N$] until the total chain length reaches $N$, with $N$ being the designated total number of aliphatic carbon atoms. Redundant generations are detected and removed via a "rotate and compare" algorithm, which rotates the molecule by $\pi/3$ each iteration and compares it with all existing molecules with the same aromatic size and aliphatic ratio.

These 1511 hydrocarbon molecules currently available in the "kerogen genome database" can be categorized into four groups based on their molecular structures: 742 molecules are poly aromatic hydrocarbon (PAH) molecules purely composed of six-member rings, ranging from 1 (benzene) to 10 rings either from the NIST online library of PAH molecules [20, 21] or generated computationally from our own random generation algorithm. 488 molecules are composed of six-ring PAH with fully saturated sp3 carbon side chains randomly attached to edges of the structures. 211 molecules are 5-carbon ring structures, also retrieved from the NIST online library. 60 PAH structures with unsaturated side chains containing C=C bonds are included to test the effects of extended conjugation with conjugated side chains.

We note that the database employed here contains molecular compounds with only hydrogen and carbon. While this limits the ability to represent the complete hetero-chemistry of natural OM that contain other heteroatoms such as oxygen, nitrogen and sulfur, the database realistically captures the molecular bonding configurations represented in the relevant portion of the Raman spectra of OM (1100 – 1700 $cm^{-1}$). We emphasize that the collective analysis approach itself is not limited by the size of the database and can be readily applied to expanded libraries containing additional molecular structures.

2.3. Calculation of Raman spectra

All the Raman spectra in this database are calculated with density functional theory as implemented in the Gaussian'09 package. We used the Perdew-Burke-Ernzerhof functional [22] to describe the exchange-correlation interaction of the electrons. Atomic orbitals are expanded with the cc-pVTZ basis set [23]. Raman activities are calculated from the differentiation of dipole derivatives with respect to the external electric field. All the molecular structures are fully relaxed using the Berny algorithm before calculating their Raman spectra. We compute Raman intensity from Raman activity following the same method used by Murphy *et. al.*[24] For visualization and comparison with experimental spectrum, discrete lines are converted to continuous Raman spectrum using Gaussian smearing of 5 wavenumbers.

2.4. Genetic Algorithm

An overview of the genetic algorithm (GA) we implemented in our approach is shown in Figure 3. At the beginning of each GA iteration, the best $N$ parent ensembles ($N$=30) are carried over directly to the child generation, increasing the chance for its "genes" (or molecular structures) to "survive" (to participate in the next generation). The rest of the child generation is filled by random mutations and crossover from the full parent generation. Here, the two operations of "mutation" and "crossover" are defined as randomly swapping one molecular fingerprint in the ensemble with another one in the database, and exchanging half of the molecular fingerprints with another ensemble, respectively. In this process the best candidate of the child generation will be no worse than the best candidate of the parent generation, and results can improve as the genetic algorithm evolves. However, the use of only iterative crossover and mutation risks locking the ensemble into a local minimum far away from better solutions. To solve this problem, we introduced the "Genesis Flood" mechanism, wherein once every thousand generations, every ensemble in the parent generation is destroyed except for the most fit, and the rest of the population is filled with random ensembles. A typical GA run consists of 50,000 iterations in order to ensure convergence. More Details and the performance of our implementation can be found in the Supplementary Information.

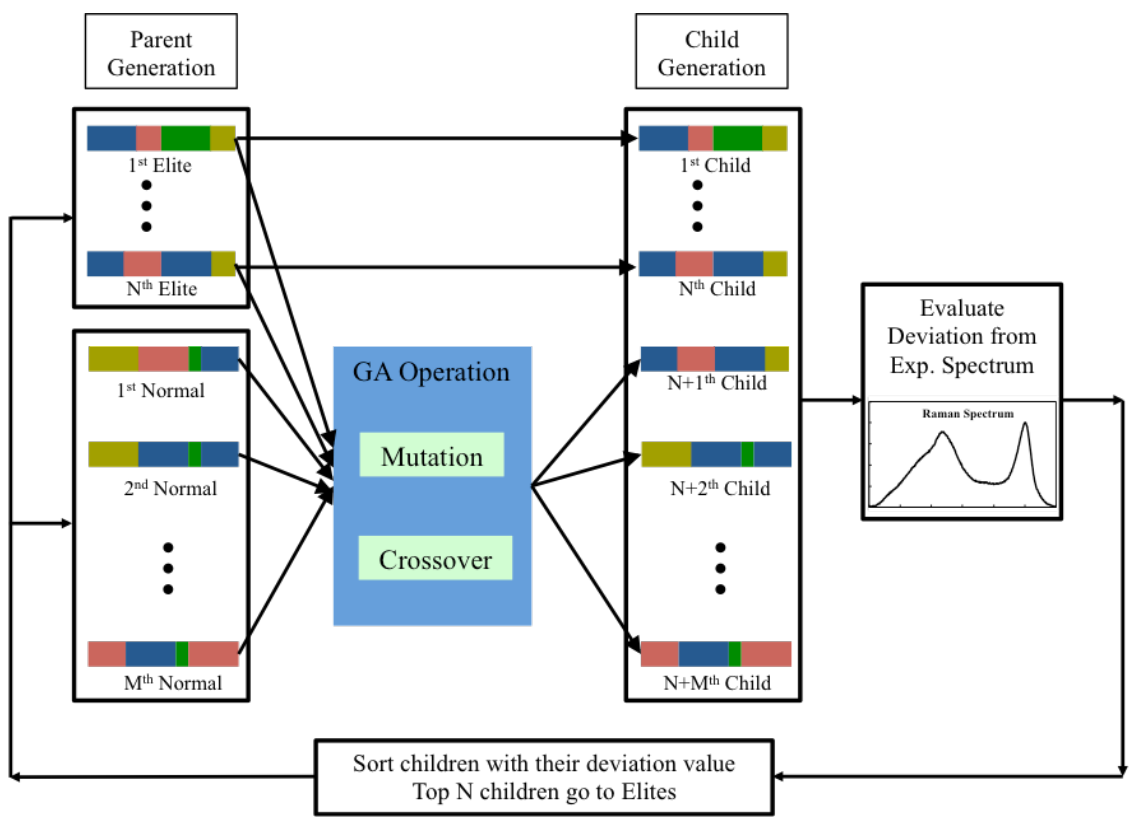

Fig. 3. Flow chart of the genetic algorithm iterations as implemented in our collective analysis code. Each colored bar represents one ensemble of molecular fingerprints in which each color block represents one molecular fingerprint. Mutation and Crossover operations are explained in the Methods section.

2.5. Raman spectroscopy

Experimental Micro-Raman spectra were acquired using a Horiba LabRAM 800 HR spectrometer equipped with a He-Ne (632.817 nm) laser as the excitation source and a Peltier-cooled CCD detector. The laser was focused on the sample with a 400 nm confocal hole using the 100X objective under reflected illumination. The laser spot on the sample was ~800 nm in diameter and had a power of ~4 mW at the sample surface. A calibrated edge high band filter (lowest wavenumber: ~70 $cm^{-1}$) was used to minimize the elastic backscattered signal. A minimum of 10 independent spots was analyzed on each sample and data were collected from 5 to 60 seconds per spot depending upon the Raman/Fluorescence intensity. The full spectral window for each acquisition is from -50 to 4000 $cm^{-1}$. To reduce the amount of artifacts introduced by the background subtraction, due to the highly non-linear background over the full spectra, the spectra was broken down into several regions of interest. The first-order spectral window for the organic region was usually taken from 1000 to 1800 $cm^{-1}$. The background subtraction in this spectral window is performed using $2^{th}$ order polynomial functions.

2.6 Organic matter sample selection

OM samples are obtained from the Argonne Premium Coal Bank [25] and Penn State Coal Bank [26]. All samples were delivered in 1 um size powder sealed under Ar. To prevent oxidation, the coals were preserved in an inert environment when not actively used for the experiments. Detailed information of all samples is listed in Table 1. While chemically heterogeneous at the molecular scale, the choice of type III kerogens, whose Raman spectra is homogenous and reproducible at the macroscale, is made to allow a meaningful validation of our approach, with available bulk scale comprehensive analytical information. As a result, the Raman spectra acquired at the microscale do not differ when acquired in different locations within the same specimen.

|  | Source | VRo | Type |
|---|---|---|---|
| OM 1 | Pittsburgh No. 8 | 0.72% | HVB |
| OM 2 | Stockton seam | 0.77% | HVB |
| OM 3 | Upper Freeport | 0.99% | MVB |
| OM 4 | Pocahontas No. 3 | 1.71% | LVB |

Table 1. Additional information for all four samples used in this study. HVB, MVB, and LVB are abbreviations for High-, Medium-, and Low-volatile bituminous, respectively. Vitrinite reflectance value (VRo) is a commonly used thermal maturity indicator, with higher VRo being geochemically more mature and *vice versa*.

## 3. Results and discussion

To evaluate this approach, we perform collective fingerprinting analysis on the Raman spectra of four well-characterized plant-based OM (type III kerogen) with ascending maturity. An ensemble size of 12 is used in our fingerprinting analysis as it allows for the main features of the experimental Raman spectrum to be well represented (see Supplementary Figure 1). For each sample, we carry out five separate GA searches and record the identified fingerprints. Fig. 4 shows one set of molecular fingerprints for each of the samples identified at the end of one collective analysis, together with the Raman responses of each molecular composition. Notably, a few key chemical quantities related to the molecular level chemistry can now be measured directly from the fingerprints obtained. Consequently, by comparing the chemical quantities measured from the molecular fingerprints against the bulk values of the corresponding sample obtained using other experimental approaches, we can directly evaluate the accuracy of our fingerprints. For instance, by counting the number of aromatic carbon atoms in each molecule, we are able to calculate directly the average aromatic cluster size. Taking OM 1 as an example, its average aromatic cluster size calculated from our analysis is 17±2 C-atom/PAH, in good agreement with the value of 15±3 C-atom/PAH estimated from carbon-13 nuclear magnetic resonance (C-NMR) spectroscopy [27]. Further, by counting the bridgehead aromatic carbon atoms in each structure, we calculate the average mole fraction of bridgehead aromatic carbons to be 0.36±0.03, in close agreement with the value of 0.31±0.06 measured from the C-NMR experiment [27]. The atomic hydrogen/carbon ratio, which is one of the most important parameters for determining the type and origin of kerogen from the Van Krevelen diagram [28], is 0.76±0.05 measured from the collective search results, in good agreement with the estimated H/C ratio of 0.77 from elemental analysis [25]. The direct comparison and excellent agreement between C-NMR and the results obtained through our genome-inspired Raman approach illustrates the effectiveness and accuracy of the fingerprinting method.

A full comparison between the chemical quantities measured from our collective analysis approach and experimental results for the samples examined in this work, are shown in Table 1. Note that the aromatic cluster sizes and H/C ratios calculated from our analysis are significantly different between different OM samples, and most of our predicted values agree well with the C-NMR experiment. OM sample 2 shows a higher aromatic cluster size, which agrees with the lower D-peak height in its Raman spectrum but differs from C-NMR experiments in the literature [27], possibly due to inhomogeneity of the sample. Our predicted H/C ratio is significantly lower for OM 4 than OM 1 or 2, in good agreement with elemental analysis experiments [25, 26]. The aromatic cluster sizes are larger for OM 4 than other samples, which is not surprising as OM 4 is more mature than other samples used in this work (see section 2.6).

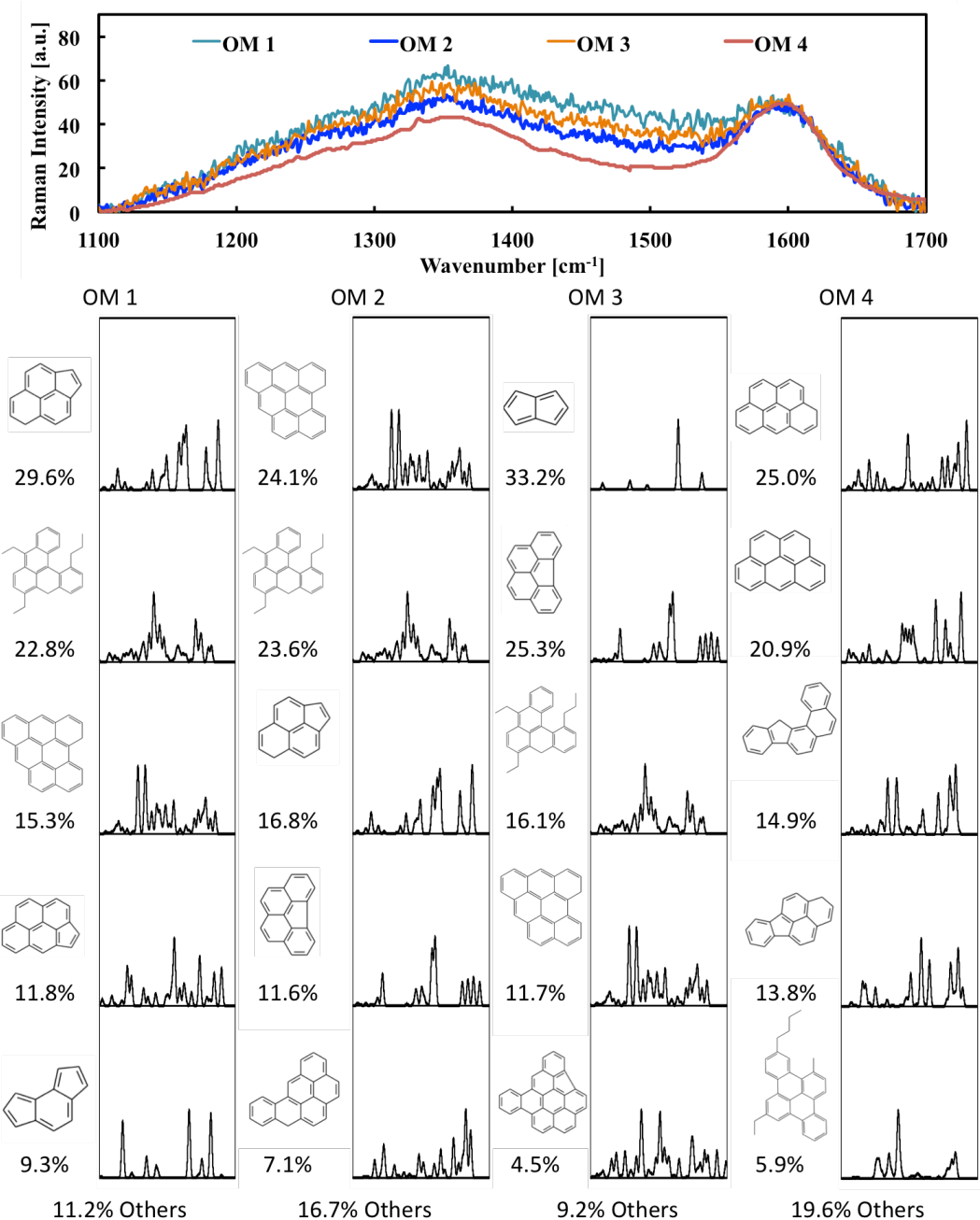

Figure. 4. List of top 5 out of 12 molecular fingerprints from one fingerprint set identified for each kerogen sample. Original Raman spectra of the four kerogen samples are shown on top. The molar ratio of each fingerprint in the 12-molecule fingerprint set is shown below the structure picture of each molecule. Raman spectrum of each molecular fingerprint in the 1100~1700 [cm$^{-1}$] wavenumber regions is shown to the right of each fingerprint.

|  | OM No. 1 | | OM No. 2 | | OM No. 3 | | OM No. 4 | |
| --- | --- | --- | --- | --- | --- | --- | --- | --- |
|  | *Cal.* | *Exp.* | *Cal.* | *Exp.* | *Cal.* | *Exp.* | *Cal.* | *Exp.* |
| H/C | 0.76±0.05 | 0.77 | 0.73±0.04 | 0.76 | 0.72±0.09 | 0.66 | 0.65±0.02 | 0.65 |
| AC | 17±2 | 15±3 | 20±1 | 14±3 | 18±2 | 18±3 | 20±2 | N/A |
| $\chi_b$ | 0.36±0.03 | 0.31±0.06 | 0.41±0.01 | 0.29±0.06 | 0.39±0.02 | 0.36±0.03 | 0.38±0.05 | N/A |

Table 2. Molecular chemistry quantities calculated from our collective fingerprinting analysis. Molar H/C ratio ("H/C"), aromatic cluster size ("AC"), and mole fraction of bridgehead aromatic carbon ("$\chi_b$") are in good agreement with the experimental values from the literature [25-27].

We also notice that among the molecular fingerprints identified for OM 4, less than 10% has an aliphatic component. In contrast, aliphatic compounds appear with a significant predicted percentage (>20%) in the fingerprints identified for OM 1 or OM 2. This comparison resonates with the fact that strong reduction of aliphatic components is a consequence of catagenesis, which leaves a heavily aromatic backbone as the major component in OM [29]. Furthermore, we have found aromatic carbon clusters with rings of five carbon atoms, which are major components in molecular representations of type III kerogens with similar maturity [30, 31]. Five-carbon rings have a higher mole ratio of bridgehead carbon atoms than six-carbon ring aromatic clusters with the same number of carbon atoms. Molecules containing five-carbon ring structures were not explicitly considered in previous C-NMR experiments [27]. One example of such molecules can be found in the fourth group in Fig. 2b. We note that although the comparison is focused on C-NMR measurements, the method developed here can be complementary to other experimental techniques that probe C-H functionality in OM (such as FT-IR) or elemental composition (mass spectrometry).

## 4. Conclusion

In this work, we present a genome-inspired method to identify from the Raman spectra the most representative molecular chemistry of complex heterogeneous carbonaceous materials. Successful application of our approach critically depends on two major components: an extensible Raman spectra database of individual molecular structures that covers the chemistry represented in the Raman spectra of the target material, and an efficient collective analysis algorithm that allows for identification of an ensemble of molecules. We show that using a "kerogen genome database" to represent the OM chemistry and using a genetic algorithm for collective analysis, it is possible to extract accurate molecular information from the OM Raman spectra: The micron-scale chemical fingerprinting can be directly compared with bulk chemical data. The genomic nature of our approach and the statistical convergence of the molecular fingerprints suggest that similarly to genetics for bio species, genetics of molecular structures can be established for OM to systematically quantify its molecular composition. The molecular structures identified through this method allows for the construction of representative molecular mixtures that can serve as the basis for large-scale atomistic simulations of OM [30-32]. More broadly, our work demonstrates the power of combining *ab initio* calculations and data-mining methods with information-rich experiments to elevate our understanding of the underlying fundamental science. While it is currently being applied to evaluate Raman spectra, the inclusion of additional computed spectral data (infrared absorption, for instance) would allow for potential application of the same collective fingerprinting approach in other characterization techniques such as FT-IR.

**Acknowledgments:** This work is supported by Shell Oil Company under the MIT Energy Initiative.

Appendix A. Supplementary data

Supplementary data associated with this article can be found in the online version.